\documentclass[11pt,a4paper,english,superscriptaddress,aps,preprint]{revtex4}
\usepackage{xcolor}
\usepackage{amsfonts}
\usepackage{graphics}
\usepackage{graphicx}
\usepackage{hyperref}
\usepackage{slashed}
\usepackage{amsmath}
\usepackage{amssymb}
\makeatletter
\usepackage{babel}
\newcommand{\bea}{\begin{eqnarray}}
\newcommand{\eea}{\end{eqnarray}}

\begin{document}

\title{Two-loop effective potential for general higher-derivative superfield models}

\author{F. S. Gama}
\email{fisicofabricio@yahoo.com.br}
\affiliation{Departamento de F\'{\i}sica, Universidade Federal da Para\'{\i}ba\\
 Caixa Postal 5008, 58051-970, Jo\~ao Pessoa, Para\'{\i}ba, Brazil}

\author{J. R. Nascimento}
\email{jroberto@fisica.ufpb.br}
\affiliation{Departamento de F\'{\i}sica, Universidade Federal da Para\'{\i}ba\\
 Caixa Postal 5008, 58051-970, Jo\~ao Pessoa, Para\'{\i}ba, Brazil}

\author{A. Yu. Petrov}
\email{petrov@fisica.ufpb.br}
\affiliation{Departamento de F\'{\i}sica, Universidade Federal da Para\'{\i}ba\\
 Caixa Postal 5008, 58051-970, Jo\~ao Pessoa, Para\'{\i}ba, Brazil}

\begin{abstract}
For a three-dimensional higher-derivative scalar superfield model, and a four-dimensional higher-derivative chiral superfield model, where the higher-derivative operators are described by polynomials of arbitrary degrees, we explicitly calculate the two-loop effective potential.
 \end{abstract}

\maketitle

\section{Introduction}

The effective potential (EP) is defined to be the zero-order term in the derivative expansion of the effective action \cite{definition,breaking}. It is used as a powerful  theoretical tool for investigating issues related to the ground state of a theory, such as false vacuum decay \cite{falsevacuum}, spontaneous symmetry breaking \cite{breaking}, and the phenomenon of symmetry restoration \cite{restoration}. On the other hand, in supersymmetric theories, it is natural to describe the low-energy effective dynamics in terms of the superfield effective potential (SEP) defined in \cite{BKY}. The reason for it is that the SEP is formulated in the superspace, so that it can be calculated entirely on the base of superfield methods, which allows us to maintain the manifest supersymmetry at all stages of the calculation. The SEP, in the one-loop and, in certain cases, two-loop approximations, has been investigated, besides of the Wess-Zumino model \cite{BKY,BKP}, in a vast variety of theories without higher derivatives, such as super-Yang-Mills theory \cite{SYM}, three-dimensional superfield theories \cite{FGLNPSS,BMS,3d}, non-renormalizable supersymmetric models \cite{NR}, and 2-form gauge superfield theories \cite{2form}.

In this work, we will study the SEP in higher-derivative theories. There are at least four reasons for studying higher-derivative theories. First, these theories emerge naturally in the low-energy limit of string models \cite{string}, in the context of effective field theory \cite{EFT}, and as counterterms necessary to renormalize semiclassical gravity theories \cite{Shapiro}. Second, the inclusion of higher-derivative terms in a field theory is also motivated by the fact that the resulting extended theory displays singularity-free classical solutions \cite{BPS}, for example, Newtonian singularities disappear in higher-derivative gravity \cite{newtonsing}, and the infinite-derivative gravity provides a solution to the Big Bang singularity problem \cite{bigbangsing}. Third, higher-derivative theories exhibit a highly improved behavior of propagator in the ultraviolet regime, this property allows us to build a finite version of QED \cite{LW} and a renormalizable gravitational theory \cite{Stelle}. Moreover, one can derive exact beta functions by working only at the one-loop level in the super-renormalizable higher-derivative gravity \cite{ALS}, and a higher-derivative extension of the standard model can solve the hierarchy problem \cite{GOW}. Fourth, the inclusion of higher-derivative terms in a field theory can also be treated as a regularization scheme \cite{Slavnov} which, when applied to supersymmetric theories, has advantages in comparison with some of the more traditional schemes, as the higher-derivative regularization does not lead to ambiguities and does not break the supersymmetry at any perturbative
order \cite{Stepanyantz}.

On the other hand, it is well-known that higher-derivative theories tend to contain extra degrees of freedom with negative energy, which leads to a unbounded Hamiltonian at both classical and quantum levels \cite{unbounded}. It is also possible to shift the problem from the lack of a ground state to the lack of unitarity, by trading the negative energy states for negative norm states (or ghosts) \cite{Woodard}. Fortunately, progress has been made in recent decades in fighting ghosts of higher-derivative theories \cite{works}. For example, the ghost degrees of freedom can be avoided if one allows an infinite number of derivatives in the action of the theory, where the infinite series in derivatives of the fields represent the expansion of an entire function \cite{nonlocal} (see Refs. \cite{susynonlocal} for supersymmetric versions of nonlocal theories). It was shown in Refs. \cite{stability} that even if the Hamiltonian is not bounded from below, there are islands of stability in some interacting higher-derivative theories in which the classical trajectories of the system do not collapse, but exhibit oscillatory behavior. Alternative to the Feynman prescription, in \cite{fakeons} was proposed a new quantization prescription for the unphysical poles of the free higher-derivative propagators. This new prescription turns the ghost degrees of freedom of the theory into fake ones (or ``fakeons"). By turning ghosts into fakeons allows us to make the higher-derivative theories unitary. In the light of all these investigations, we will take the view that higher-derivative theories are viable physical theories.

The encouraging properties mentioned above motivated studies of the SEP in higher-derivative models within different contexts \cite{SHD,GGNPS}. For example, one-loop corrections to the EP were explicitly computed for different three-dimensional higher-derivative superfield theories in \cite{GNP,GNP2}. In Ref. \cite{CMP}, the decoupling of heavy chiral superfields in supersymmetric theories with higher-derivative terms was investigated with use of the one-loop EP. Recently, we formulated supersymmetric gauge theories with higher derivatives in the matter sector and calculated the one-loop SEP for such theories \cite{GNP3}. On the other hand, two-loop contributions to the EP in higher-derivative superfield theories never have been studied. Thus, in this paper, we intend to go beyond the one-loop approximation and calculate the two-loop SEP for two higher-derivative models: the first is a three-dimensional higher-derivative scalar superfield model, and the other one is a four-dimensional higher-derivative chiral superfield model. These models are quite general due to the fact that the higher-derivative operators are described by polynomials  of arbitrary degrees. It is worth mentioning that we have already calculated the one-loop correction to the SEP of these models in our previous works \cite{GNP,DGNP}, but our final results were given in terms of integrals over the momenta. On the other hand, we will obtain explicit expressions for the SEPs in the present paper. Therefore, our present manuscript   includes a  further development of studies performed in our previous works \cite{GNP,DGNP}. 

This paper is structured as follows. In section II, we calculate the one- and two-loop contributions to the SEP in the higher-derivative scalar superfield model. In section III, we calculate the one- and two-loop contributions to the SEP in the higher-derivative chiral superfield model. In section IV, we draw our conclusion from the results obtained and suggest a possible continuation of this study. In this work, we will limit ourselves to the Euclidean space-time which is more convenient for treating higher derivative theories within need to solve the ghost problem \cite{TrodFont}.

\section{Higher-Derivative Scalar Superfield Model}

In this section, we start with a higher-derivative extension of the self-interacting scalar superfield model in the $\mathcal{N}=1, d=3$ superspace:
\bea
\label{3dHM}
S[\Phi]=\frac{1}{2}\int d^5z\Phi h(D^2)\Phi+\int d^5zV(\Phi) \ ,
\eea
where $V(\Phi)$ is an analytic function of the real scalar superfield $\Phi(z)$. The operator $h(D^2)$ is assumed
to be a polynomial function of $D^2$. Thus, this operator can be expressed as
\bea
h(D^2)=\sum_{i=1}^N h_i(D^2)^i \ . \ 
\eea
In order to recover the standard model for the real scalar multiplet \cite{GGRS}, we have to make the assumption that $h_1=1$ and the coefficients $h_2,\ldots,h_N$ vanish in some suitable limit. It is worth to point out that, with the help of the identity $(D^2)^2=\Box$, it is possible to write $h(D^2)=g(\Box)+f(\Box)D^2$, where $g(\Box)$ and $f(\Box)$ are some polynomial functions. Accordingly, the model (\ref{3dHM}) is the same as the one discussed in \cite{GNP}.

In the present section, we are interested in calculating the SEP $V_{eff}$ of the model (\ref{3dHM}) in the two-loop approximation. In the $\mathcal{N}=1, d=3$ superspace, the SEP is defined in a manner similar to the EP in the standard quantum field theory \cite{BOS}, namely, the $V_{eff}$ is the zero-order term in the local approximation for the superfield effective action:
\bea
\label{3ddefinition}
\Gamma[\Phi]=\int d^5z\left[V_{eff}(\Phi)+\cdots\right] \ ,
\eea
where the dots denote the terms that depend on covariant derivatives.

The EP is calculated within the framework of the loop expansion
\bea
\label{3ddefeffpot}
V_{eff}(\Phi)=V(\Phi)+V^{(1)}(\Phi)+V^{(2)}(\Phi)+\cdots \ ,
\eea
where $V^{(1)}$ and $V^{(2)}$ represent the one- and two-loop quantum corrections to the tree-level potential $V(\Phi)$, respectively.

Now, let us begin with the calculation of the one- and two-loop corrections $V^{(1)}$ and $V^{(2)}$. In this work, the EP is obtained with use of the background field method \cite{DeWitt}. Following this method, we first make the background-quantum splitting $\Phi\rightarrow\Phi+\phi$ in (\ref{3dHM}), where $\Phi$ is now the background field, and $\phi$ is the quantum one. Then, we expand $S[\Phi+\phi]$ around the background superfield $\Phi$ and disregard terms proportional to the derivatives of $\Phi$, as required for obtaining the effective potential. In particular, for computing the one-loop correction $V^{(1)}$, we have to expand $S[\Phi+\phi]$ and keep only quadratic terms in the quantum superfield $\phi$. Therefore, it follows from (\ref{3dHM}) that
\bea
\label{3ds2}
S[\Phi+\phi]=\frac{1}{2}\int d^5z\phi\left[h(D^2)+V^{\prime\prime}\right]\phi \ .
\eea
By formal integrating over the quantum superfield in (\ref{3ds2}), we obtain the following one-loop contribution to the effective action:
\bea
\label{3dgamma}
\Gamma^{(1)}=-\frac{1}{2}\textrm{Tr}\ln\left[h(D^2)+V^{\prime\prime}\right] \ .
\eea
The fundamental theorem of algebra implies that every polynomial of degree $N$ has, taking into account the multiplicity, exactly $N$ complex roots. For the sake of simplicity, we assume that the coefficients of the polynomial $P(z)\equiv h(z)+V^{\prime\prime}$ are chosen in such a way that all its roots are distinct and real. Thus, this polynomial can be factorized as a product of primitive multipliers:
\bea
\label{3dpolynomial}
P(z)\equiv h(z)+V^{\prime\prime}=\sum_{k=0}^N h_kz^k=h_N\prod_{k=1}^N\left(z+m_k\right) \ ,
\eea
where $h_0\equiv V^{\prime\prime}$, while $-m_k$ are the roots of $P(z)$. Accordingly, Eq. (\ref{3dgamma}) can be rewritten as
\bea
\label{3dfactored}
\Gamma^{(1)}=-\frac{1}{2}\textrm{Tr}\ln h_N-\frac{1}{2}\sum_{k=1}^N\textrm{Tr}\ln\left(D^2+m_k\right) \ .
\eea
Note that the first trace does not depend on the background superfield, then we can discard it. The typical contribution to second trace, that is, $\textrm{Tr}\ln\left(D^2+m\right)$ has already been evaluated in Ref. \cite{FGLNPSS}. Thus, (\ref{3dfactored}) yields
\bea
\Gamma^{(1)}=-\frac{1}{16\pi}\int d^5z\sum_{k=1}^N m^2_k \ ,
\eea
It is convenient to express this result in terms of the coefficients of the polynomial $P(z)$ (\ref{3dpolynomial}). Thus,
\bea
\label{3dgamma_2}
\Gamma^{(1)}&=&-\frac{1}{16\pi}\int d^5z\left[\left(\sum_{k=1}^N m_k\right)^2-2\sum_{1\leq k<l\leq N} m_k m_l\right]\nonumber \\
&=&-\frac{1}{16\pi}\int d^5z\left[\left(\frac{h_{N-1}}{h_N}\right)^2-2\frac{h_{N-2}}{h_N}\right] \ .
\eea
 where we have used Vieta's formulas
\bea
\sum_{1\leq i_1<i_2<\cdots<i_k\leq N} m_{i_1}m_{i_2}\cdots m_{i_k}=\frac{h_{N-k}}{h_N} \ .
\eea
Since only the coefficient $h_0=V^{\prime\prime}$ depends on the background superfield, then the only nontrivial correction is the one with $N=2$. Therefore, it follows from (\ref{3dgamma_2}) and (\ref{3ddefinition}) that the one-loop EP is given by
\bea
\label{3dfinal1loop}
V^{(1)}(\Phi)=\left\{\begin{array}{rcl}
\displaystyle \frac{h_2^{-1}}{8\pi}V^{\prime\prime}, & \displaystyle \mbox{if} & \displaystyle N=2;\\
\displaystyle 0 \ \ \ \ , & \displaystyle \mbox{if} & \displaystyle N>2.
\end{array}
\right. 
\eea
Now, let us begin with studying the two-loop correction to the EP $V^{(2)}$. In this case, the action $S[\Phi+\phi]$ must be expanded in a series in $\phi$, and we must keep the terms up to the fourth order. Therefore, we obtain
\bea
\label{3ds4}
S[\Phi+\phi]=\frac{1}{2}\int d^5z\phi\left[h(D^2)+V^{\prime\prime}\right]\phi+\int d^5z\left[\frac{1}{3!}V^{\prime\prime\prime}\phi^3+\frac{1}{4!}V^{\prime\prime\prime\prime}\phi^4\right] \ .
\eea
The $3$- and $4$-point vertex functions can be obtained directly from (\ref{3ds4}). They are 
\bea
\label{3d3vertex}
\frac{\delta^3 S}{\delta\phi(z_1)\delta\phi(z_2)\delta\phi(z_3)}&=&V^{\prime\prime\prime}(z_3)\delta^5(z_1-z_2)\delta^5(z_2-z_3) \ ;\\
\label{3d4vertex}
\frac{\delta^4 S}{\delta\phi(z_1)\delta\phi(z_2)\delta\phi(z_3)\delta\phi(z_4)}&=&V^{\prime\prime\prime\prime}(z_4)\delta^5(z_1-z_2)\delta^5(z_2-z_3)\delta^5(z_3-z_4) \ .
\eea
Moreover, we also obtain from (\ref{3ds4}) the following propagator
\bea
G(z,w)=-\frac{1}{h(D^2)+V^{\prime\prime}}\delta^5(z-w)=-\frac{h_N^{-1}}{\prod_{k=1}^{N}\left(D^2+m_k\right)}\delta^5(z-w) \ ,
\eea
where we have used (\ref{3dpolynomial}).

\begin{figure}[!h]
\begin{center}
\includegraphics[angle=0,scale=0.60]{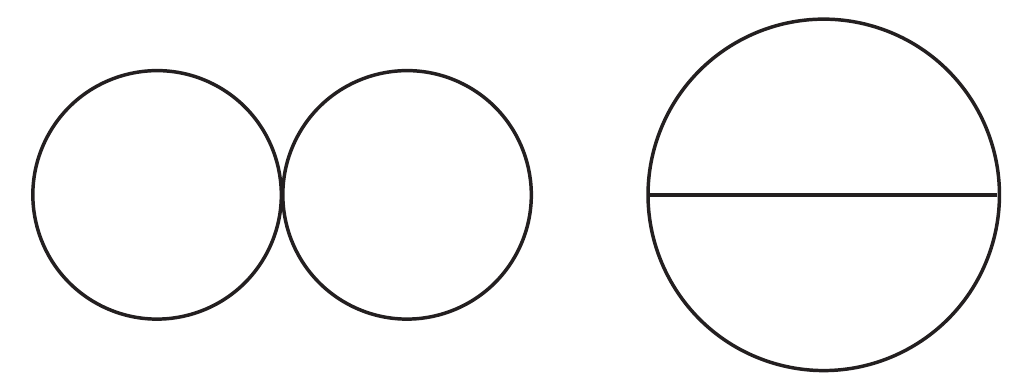}
\end{center}
\caption{Topologies of the supergraphs for the two-loop contributions to the effective potential.}
\label{fig:two-loop_supergraphs}
\end{figure}

In order to present our results as Feynman integrals with known analytical solutions, we want to express $G(z,w)$ in terms of simple propagators. To do this, we notice that $\left[h(z)+V^{\prime\prime}\right]^{-1}$ has $N$ simple poles at $z=-m_k$, for $k=1,\ldots,N$. This allows us to find, by means of theory of residues \cite{MH}, the following so-called partial fraction representation of $G(z,w)$: 
\bea
\label{3dpfr}
G(z,w)=-h_N^{-1}\sum_{k=1}^N\frac{c_k}{D^2+m_k}\delta^5(z-w)=-h_N^{-1}\sum_{k=1}^N c_k\frac{D^2-m_k}{\Box-m^2_k}\delta^5(z-w) \ ,
\eea
where $c_k$'s are the residues of $\left[h(z)+V^{\prime\prime}\right]^{-1}$ at $z=-m_k$. These residues are given by
\bea
\label{3dresidues}
c_k=\prod_{j\neq k}\frac{1}{m_j-m_k} \ .
\eea
With the propagator and vertex functions in hand, we now turn to computing the two-loop supergraphs which contribute to $V^{(2)}(\Phi)$. Such supergraphs have the topologies of the "$\infty$" and "$\ominus$" (see Fig. \ref{fig:two-loop_supergraphs}) and correspond to the following contributions
\bea
\label{3dinfty}
\Gamma^{(2)}_\infty &=&\frac{1}{8}\int d^5z_1d^5z_2d^5z_3d^5z_4\frac{\delta^4 S}{\delta\phi(z_1)\delta\phi(z_2)\delta\phi(z_3)\delta\phi(z_4)}G(z_1,z_2)G(z_3,z_4) \ ;\\
\label{3dominus}
\Gamma^{(2)}_\ominus &=&\frac{1}{12}\int d^5z_1d^5z_2d^5z_3d^5z_4d^5z_5d^5z_6\frac{\delta^3 S}{\delta\phi(z_1)\delta\phi(z_2)\delta\phi(z_3)}\frac{\delta^3 S}{\delta\phi(z_4)\delta\phi(z_5)\delta\phi(z_6)}\nonumber\\
&\times& G(z_1,z_4)G(z_2,z_5)G(z_3,z_6) \ .
\eea
Let us first consider the contribution $\Gamma^{(2)}_\infty$. Substituting (\ref{3d4vertex}) into (\ref{3dinfty}) and using the delta functions to evaluate three integrals, we find
\bea
\label{3dinfty_2}
\Gamma^{(2)}_\infty &=&\frac{1}{8}\int d^5zV^{\prime\prime\prime\prime}(z)G^2(z,z) \ .
\eea
Next, we use the partial fraction representation (\ref{3dpfr}) to calculate $G(z,z)$. Thus, we have 
\bea
\label{gzz}
G(z,z)&=&\left.-h_N^{-1}\sum_{k=1}^N c_k\frac{D^2-m_k}{\Box-m^2_k}\delta^5(z-z^\prime)\right|_{z=z^\prime}=\left.-h_N^{-1}\sum_{k=1}^N c_k\frac{1}{\Box-m^2_k}\delta^3(x-x^\prime)\right|_{x=x^\prime}\nonumber\\
&=&-\frac{h_N^{-1}}{4\pi}\sum_{k=1}^N c_km_k \ ,
\eea
where, to obtain the last equality, we have used the Fourier representation of the delta function and calculated the Feynman integral. 

It is worth to point out that the following identity holds (although we do not know its general proof, it was checked with use of Mathematica to be valid at least up to $N=10$ which allows to conclude that it is apparently valid for any $N$):
\bea
\label{3didentity}
\sum_{k=1}^N c_km_k=\left\{\begin{array}{rcl}
\displaystyle -1, & \displaystyle \mbox{if} & \displaystyle N=2;\\
\displaystyle 0, & \displaystyle \mbox{if} & \displaystyle N>2.
\end{array}
\right. 
\eea
Therefore, substituting (\ref{gzz}) into (\ref{3dinfty_2}) and using the identity (\ref{3didentity}), we can show that the first two-loop contribution to the EP is given by
\bea
\label{3dinfty2loop}
V^{(2)}_\infty(\Phi)=\left\{\begin{array}{rcl}
\displaystyle \frac{h_2^{-2}}{128\pi^2}V^{\prime\prime\prime\prime}, & \displaystyle \mbox{if} & \displaystyle N=2;\\
\displaystyle 0 \ \ \ \ \ \ \ , & \displaystyle \mbox{if} & \displaystyle N>2.
\end{array}
\right. 
\eea
Note that, similarly to the one-loop contribution (\ref{3dfinal1loop}), this one is also trivial for $N>2$.

Now, let us proceed the calculation of $\Gamma^{(2)}_\ominus$. After substituting (\ref{3d3vertex}) into (\ref{3dominus}), we can use the delta functions to perform the integrals and write down the following equivalent expression:
\bea
\label{3dominus_2}
\Gamma^{(2)}_\ominus&=&\frac{1}{12}\int d^5z_1d^5z_4V^{\prime\prime\prime}(z_1)V^{\prime\prime\prime}(z_4)G^3(z_1,z_4)\nonumber\\
&=&\frac{1}{12}\int d^5z_1d^5z_4V^{\prime\prime\prime}(z_1)V^{\prime\prime\prime}(z_4)\left[-h_N^{-1}\sum_{k=1}^N c_k\frac{D^2_1-m_k}{\Box_1-m^2_k}\delta^5(z_1-z_4)\right]G^2(z_1,z_4) \ .
\eea
By using the properties of the spinor derivatives and the delta functions, it is possible to show that
\bea
&&\delta^2(\theta_1-\theta_4)G(z_1,z_4)=-h_N^{-1}\sum_{k=1}^N\frac{c_k}{\Box_1-m^2_k}\delta^5(z_1-z_4) \ ;\\
&&\delta^2(\theta_1-\theta_4)D_{1\alpha}G(z_1,z_4)=0 \ ;\\
&&\delta^2(\theta_1-\theta_4)D_1^2G(z_1,z_4)=h_N^{-1}\sum_{k=1}^N\frac{c_km_k}{\Box_1-m^2_k}\delta^5(z_1-z_4) \ .
\eea
Thus, after doing the $D$-algebra in (\ref{3dominus_2}) and using the identities above, we arrive at
\bea
\Gamma^{(2)}_\ominus&=&\frac{h_N^{-3}}{4}\int d^2\theta_1d^3x_1d^3x_4\left(V^{\prime\prime\prime}\right)^2\sum_{k=1}^N\sum_{l=1}^N\sum_{m=1}^Nc_kc_lc_mm_m\left[\frac{1}{\Box_1-m^2_k}\delta^3(x_1-x_4)\right]\nonumber\\&\times&\left[\frac{1}{\Box_1-m^2_l}\delta^3(x_1-x_4)\right]\left[\frac{1}{\Box_1-m^2_m}\delta^3(x_1-x_4)\right]\nonumber\\
&=&-\frac{h_N^{-3}}{4}\int d^5z\left(V^{\prime\prime\prime}\right)^2\sum_{k=1}^N\sum_{l=1}^N\sum_{m=1}^Nc_kc_lc_mm_m\int \frac{d^3pd^3q}{(2\pi)^6}\frac{1}{p^2+m^2_k}\frac{1}{q^2+m^2_l}\nonumber\\
&\times&\frac{1}{(p+q)^2+m^2_m} \ ,
\eea
where we have used the Fourier transform of the bosonic delta functions. 

By performing these two-loop integrals using the formula given in Ref. \cite{FKRS} (see \cite{TTH} for a detailed derivation), we obtain 
\bea
\label{3dominus_3}
\Gamma^{(2)}_\ominus=-\frac{h_N^{-3}}{128\pi^2}\int d^5z\left(V^{\prime\prime\prime}\right)^2\sum_{k=1}^N\sum_{l=1}^N\sum_{m=1}^Nc_kc_lc_mm_m\left\{\frac{1}{2\varepsilon}+1-\ln\left[\frac{(m_k+m_l+m_m)^2}{\bar{\mu}^2}\right]\right\} \ ,
\eea
where $\varepsilon=\frac{1}{2}(3-d)$ and $\bar{\mu}$ is an arbitrary mass parameter. Note that each Feynman integral above gives rise to a singularity $\varepsilon^{-1}$. However, it is worth
pointing out that these singularities are spurious and cancel out due to the identity
\bea
\sum_{k=1}^Nc_k=0 \ ,
\eea
which follows from (\ref{3dresidues}) and can be checked with use of Mathematica.

Therefore, the second two-loop contribution to the EP, evaluated with application of (\ref{3dominus_3}), is equal to:
\bea
\label{3dominus2loop}
V^{(2)}_\ominus(\Phi)=\frac{h_N^{-3}}{128\pi^2}\left(V^{\prime\prime\prime}\right)^2\sum_{k=1}^N\sum_{l=1}^N\sum_{m=1}^Nc_kc_lc_mm_m\ln\left[\frac{(m_k+m_l+m_m)^2}{\bar{\mu}^2}\right] \ .
\eea
Finally, by substituting the results (\ref{3dfinal1loop}), (\ref{3dinfty2loop}), and (\ref{3dominus2loop})
 into (\ref{3ddefeffpot}) we arrive at the full SEP
\bea
\label{3dfull}
V_{eff}(\Phi)&=&V+\frac{h_2^{-1}}{8\pi}V^{\prime\prime}+\frac{h_2^{-2}}{128\pi^2}V^{\prime\prime\prime\prime}+\frac{h_2^{-3}}{128\pi^2}\left(V^{\prime\prime\prime}\right)^2\nonumber\\
&\times&\sum_{k=1}^2\sum_{l=1}^2\sum_{m=1}^2c_kc_lc_mm_m\ln\left[\frac{(m_k+m_l+m_m)^2}{\bar{\mu}^2}\right] \ ,
\eea
for $N=2$, and
\bea
\label{3dfull_2}
V_{eff}(\Phi)=\frac{h_N^{-3}}{128\pi^2}\left(V^{\prime\prime\prime}\right)^2\sum_{k=1}^N\sum_{l=1}^N\sum_{m=1}^Nc_kc_lc_mm_m\ln\left[\frac{(m_k+m_l+m_m)^2}{\bar{\mu}^2}\right] \ ,
\eea
for $N>2$.

In contrast to EP in the scalar superfield model without higher derivatives \cite{FGLNPSS}, we note that our main result is finite and do not need any renormalization. Additionally, our result is valid for any analytical function $V(\Phi)$. We note that this contribution displays the logarithmic behavior despite vanishing of divergences. The similar situation was earlier shown to occur in a certain four-dimensional chiral superfield model \cite{GGNPS}.

In order to illustrate the methodology developed in this section, let us now consider an explicit example. Let us take
\bea
h(D^2)=\frac{\Box}{\Lambda}+D^2 \ ,
\eea
where $h_2=\Lambda^{-1}>0$. In this case, we must find the roots of the quadratic function $h(z)+V^{\prime\prime}=\Lambda^{-1}z^2+z+V^{\prime\prime}$ (see Eq. (\ref{3dpolynomial})). The roots are $-m_\pm$, where
\bea
\label{3droots}
m_{\pm}=\frac{1}{2}\left(\Lambda\pm\sqrt{\Lambda(\Lambda-4V^{\prime\prime})}\right) \ .
\eea
We assume that $\Lambda>4V^{\prime\prime}$ to ensure that these roots are distinct and real.

Inserting (\ref{3droots}) into (\ref{3dresidues}), we get the residues
\bea
\label{3dresidues_2}
c_-=-c_+=\frac{1}{\sqrt{\Lambda(\Lambda-4V^{\prime\prime})}} \ .
\eea
Therefore, substituting (\ref{3droots}) and (\ref{3dresidues_2}) into (\ref{3dfull}), we obtain the full EP
\bea
V_{eff}(\Phi)&=&V+\frac{\Lambda}{8\pi}V^{\prime\prime}+\frac{\Lambda^{2}}{128\pi^2}V^{\prime\prime\prime\prime}-\frac{\Lambda^3}{256\pi^2}\frac{\left(V^{\prime\prime\prime}\right)^2}{\sqrt{\Delta^3}}\Bigg\{\big(\Lambda+\sqrt{\Delta}\big)\ln\left[\frac{9}{4}\frac{\big(\Lambda+\sqrt{\Delta}\big)^2}{\bar{\mu}^2}\right]\nonumber\\
&-&\big(3\Lambda+\sqrt{\Delta}\big)\ln\left[\frac{\big(3\Lambda+\sqrt{\Delta}\big)^2}{4\bar{\mu}^2}\right]+\big(3\Lambda-\sqrt{\Delta}\big)\ln\left[\frac{\big(3\Lambda-\sqrt{\Delta}\big)^2}{4\bar{\mu}^2}\right]-\big(\Lambda-\sqrt{\Delta}\big)\nonumber\\
&\times&\ln\left[\frac{9}{4}\frac{\big(\Lambda-\sqrt{\Delta}\big)^2}{\bar{\mu}^2}\right]\Bigg\} \ ,
\eea
where $\Delta=\Lambda(\Lambda-4V^{\prime\prime})$. So, we conclude that the two-loop effective potential is described by the logarithmic function.

\section{Higher-Derivative Chiral Superfield Model}

In the present section, we will see that the approach used in the previous section to calculate the SEP can be applied, with minor modifications, within the context of the higher-derivative chiral superfield model in the $\mathcal{N}=1, d=4$ superspace.

We start our study with the model
\bea
\label{4dHM}
S[\Phi,\bar{\Phi}]=\int d^8z\bar{\Phi}f(\Box)\Phi+\left\{\int d^6z\left[\frac{1}{2}\Phi g(\Box)\Phi+W(\Phi)\right]+h.c.\right\} \ ,
\eea
where $W(\Phi)$ is an analytic function of the chiral superfield $\Phi(z)$. The operators $f(\Box)$ and $g(\Box)$ are polynomial functions of $\Box$ defined as
\bea
\label{fg}
f(\Box)=\sum_{i=0}^{N_f} f_i\Box^i \ ; \ g(\Box)=\sum_{i=1}^{N_g} g_i\Box^i \ .
\eea
In order to match the superfield Wess-Zumino model in the low-energy limit, we assume that $f_0=1$, while $f_i$ and $g_i$ for $i\geq1$ vanish in this limit, which in the momentum space corresponds to zero momenta, $p\to 0$. The particular case of this model, with $f(\Box)=1$ and $g(\Box)\propto\Box$, has been considered in the one-loop approximation in \cite{GGNPS}.
It is worth mentioning that one-loop EP of (\ref{4dHM}) has been calculated also for different nonlocal choices of $f(\Box)$ and $g(\Box)$ in \cite{DGNP}.

By assuming that the superfield $\Phi(z)$ varies slowly in the $\mathcal{N}=1, d=4$ superspace, which is the usual condition for the K\"{a}hlerian effective potential, the effective action $\Gamma[\Phi,\bar{\Phi}]$ can be the expanded in series in covariant derivatives and written as \cite{BK,BKY}
\bea
\Gamma[\Phi,\bar{\Phi}]=\int d^8zK_{eff}(\Phi,\bar{\Phi})+\left[\int d^6zW_{eff}(\Phi)+h.c.\right]+\cdots \ .
\eea
Thus, the zero-order term in this expansion is determined by two objects, namely, the K\"{a}hlerian EP $K_{eff}(\Phi,\bar{\Phi})$ and the chiral EP $W_{eff}(\Phi)$. In this section, for the sake of simplicity, we are investigating only the K\"{a}hlerian EP (however, one can argue that the one-loop contribution to the chiral EP vanishes in this theory in a whole analogy with simpler higher-derivative models of a chiral superfield \cite{SHD}).

The K\"{a}hlerian EP admits the following loop expansion:
\bea
\label{4ddefeffpot}
K_{eff}(\Phi,\bar{\Phi})=\bar{\Phi}\Phi+K^{(1)}(\Phi,\bar{\Phi})+K^{(2)}(\Phi,\bar{\Phi})+\cdots \ ,
\eea
where $K^{(1)}$ and $K^{(2)}$ denote the one- and two-loop quantum corrections, respectively.

Let us first determine $K^{(1)}$, which is the simplest and usually most important contribution. Similar to what we did in the previous section, we make a split in the superfield $\Phi\rightarrow\Phi+\phi$ together with the analogous split for $\bar{\Phi}$. Then, we expand $S[\Phi+\phi]$ around the background superfields $\Phi$ and $\bar{\Phi}$ keeping only the quadratic terms in the quantum superfields. Therefore, it follows from (\ref{4dHM}) that
\bea
\label{4ds2}
S[\Phi+\phi,\bar{\Phi}+\bar{\phi}]=\int d^8z\bar{\phi}f(\Box)\phi+\left\{\int d^6z\frac{1}{2}\phi\left[g(\Box)+W^{\prime\prime}\right]\phi+h.c.\right\} \ ,
\eea
where the irrelevant terms involving covariant derivatives of the background (anti)chiral superfields were dropped.

It follows from (\ref{4ds2}) that the one-loop contribution to the effective action is given by the standard functional trace
\bea
\label{4dgamma}
\Gamma^{(1)}=-\frac{1}{2}\textrm{Tr}\ln\left(\begin{array}{cc}
\displaystyle g(\Box)+W^{\prime\prime} & \displaystyle f(\Box)\bar{D}^2\\
\displaystyle f(\Box)D^2 & \displaystyle g(\Box)+\bar{W}^{\prime\prime}
\end{array}\right) \ . 
\eea
We want to rewrite (\ref{4dgamma}) in terms of a some scalar operator in the same way as we did with Eq. (\ref{3dgamma}). To do this, Eq. (\ref{3dfactored}) needs to be put in a more convenient form. Thus, let us rewrite (\ref{4dgamma}) as
\bea
\label{4dgamma_2}
\Gamma^{(1)}=-\frac{1}{2}\textrm{Tr}\ln H_0-\frac{1}{2}\textrm{Tr}\ln\left(I_2+H_\Phi\right) \ ,
\eea
where
\bea
\label{h0}
H_0&\equiv&\left(\begin{array}{cc}
\displaystyle 0 & \displaystyle f(\Box)\bar{D}^2\\
\displaystyle f(\Box)D^2 & \displaystyle 0
\end{array}\right) \ ;\\
\label{hPhi}
H_\Phi&\equiv&\frac{1}{\Box f(\Box)}\left(\begin{array}{cc}
\displaystyle 0 & \displaystyle \left[g(\Box)+\bar{W}^{\prime\prime}\right]\bar{D}^2\\
\displaystyle \left[g(\Box)+W^{\prime\prime}\right]D^2 & \displaystyle 0
\end{array}\right) \ .
\eea
On the one hand, using the identity $-\frac{1}{2}\textrm{Tr}\ln H_0=-\frac{1}{4}\textrm{Tr}\ln H_0^2$ and (\ref{h0}) we are able to show that
\bea
\label{first}
-\frac{1}{2}\textrm{Tr}\ln H_0=-\frac{1}{4}\textrm{Tr}\ln\left(\begin{array}{cc}
\displaystyle \Box f^2(\Box) & \displaystyle 0\\
\displaystyle 0 & \displaystyle \Box f^2(\Box)
\end{array}\right)=-\frac{1}{4}\textrm{Tr}_+\ln\Box f^2(\Box)+h.c. \ ,
\eea
where $\textrm{Tr}_+$ designates the trace over the chiral subspace \cite{BK}.

On the other hand, because of the trace and the anti-diagonal nature of the matrix $H_\Phi$, it follows that only even powers survive in the expansion of $-\frac{1}{2}\textrm{Tr}\ln\left(I_2+H_\Phi\right)$. Indeed, it is possible to show that
\bea
-\frac{1}{2}\textrm{Tr}\ln\left(I_2+H_\Phi\right)=-\frac{1}{4}\textrm{Tr}\ln\left(I_2-H_\Phi^2\right) \ .
\eea
Thus, substituting (\ref{hPhi}) into this equation, we obtain
\bea
\label{second}
-\frac{1}{2}\textrm{Tr}\ln\left(I_2+H_\Phi\right)&=&-\frac{1}{4}\textrm{Tr}\ln\left(\begin{array}{cc}
\displaystyle 1-\frac{\left|g(\Box)+W^{\prime\prime}\right|^2}{\Box f^2(\Box)} & \displaystyle 0\\
\displaystyle 0 & \displaystyle 1-\frac{\left|g(\Box)+W^{\prime\prime}\right|^2}{\Box f^2(\Box)}
\end{array}\right)\nonumber\\
&=&-\frac{1}{4}\textrm{Tr}_+\ln\left[1-\frac{\left|g(\Box)+W^{\prime\prime}\right|^2}{\Box f^2(\Box)}\right]+h.c. \ .
\eea
Finally, substituting (\ref{first}) and (\ref{second}) into (\ref{4dgamma_2}), we find
\bea
\label{4dgamma_3}
\Gamma^{(1)}=-\frac{1}{4}\textrm{Tr}_+\ln\left[\Box f^2(\Box)-\left|g(\Box)+W^{\prime\prime}\right|^2\right]+h.c. \ .
\eea
Now we are ready to decompose the operator in the argument of the logarithm function. Note that the fundamental theorem of algebra allows us to factor the polynomial
\bea
\label{4dpolynomial}
zf^2(z)-\left|g(z)+W^{\prime\prime}\right|^2\equiv\sum_{k=0}^N a_kz^k=a_N\prod_{k=1}^N\left(z-m^2_k\right) \ ,
\eea
where again we assume that the coefficients of $zf^2(z)-\left|g(z)+W^{\prime\prime}\right|^2$ are chosen in such way that it has $N$ distinct real roots $m_k^2$. Consequently, Eq. (\ref{4dgamma_3}) can be rewritten as
\bea
\label{4dfactored}
\Gamma^{(1)}=-\frac{1}{4}\textrm{Tr}_+\ln a_N-\frac{1}{4}\sum_{k=1}^N\textrm{Tr}_+\ln\left(\Box-m^2_k\right)+h.c. \ .
\eea
The first trace does not depend on the background superfields, then we can drop it from our analysis. The second trace can be performed using the same procedure as in \cite{BMS}. Therefore, the result is given by
\bea
\Gamma^{(1)}=\frac{1}{32\pi^2}\int d^8z\sum_{k=1}^N m^2_k\left[\frac{1}{\varepsilon}+2-\ln\left(\frac{m^2_k}{\bar{\mu}^2}\right)\right] \ .
\eea
This contribution appears to be divergent. However, actually the above divergences are annihilated by the Grassmann integration. In order to see why, let us consider the Vieta's formulas
\bea
\sum_{1\leq i_1<i_2<\cdots<i_k\leq N} m_{i_1}^2m_{i_2}^2\cdots m_{i_k}^2=(-1)^k\frac{a_{N-k}}{a_N}
\eea
to rewrite the divergent part in terms of the coefficients of polynomial (\ref{4dpolynomial}). Thus, we obtain
\bea
\label{4dfactored_2}
\Gamma^{(1)}=-\frac{1}{32\pi^2}\int d^8z\left(\frac{a_{N-1}}{a_N}\right)\left(\frac{1}{\varepsilon}+2\right)-\frac{1}{32\pi^2}\int d^8z\sum_{k=1}^N m^2_k\ln\left(\frac{m^2_k}{\bar{\mu}^2}\right) \ .
\eea
Now, given the definitions (\ref{fg}) and (\ref{4dpolynomial}), we are able to state that $a_0=-|W^{\prime\prime}|^2$. Moreover, for $N_g\geq k>0$, the coefficients $a_k$ depend linearly on $W^{\prime\prime}+\bar{W}^{\prime\prime}$. Finally, for $k>N_g$, the coefficients $a_k$ do not depend on the background superfields. Thus, due to properties of the integral over the Grassmann variables, the divergent part of $\Gamma^{(1)}$ vanishes. Therefore, we can infer from (\ref{4dfactored_2}) that the one-loop K\"{a}hlerian EP is
\bea
\label{4dfinal1loop}
K^{(1)}(\Phi,\bar{\Phi})=-\frac{1}{32\pi^2}\sum_{k=1}^N m^2_k\ln\left(\frac{m^2_k}{\bar{\mu}^2}\right) \ .
\eea
We note that the result displays a logarithmic behavior despite its explicit finiteness. The same situation has been observed earlier in \cite{GGNPS} where a particular case of this theory has been considered.

Let us now consider the calculation of the two-loop effective K\"{a}hlerian potential. In the two-loop approximation, the action $S[\Phi+\phi,\bar{\Phi}+\bar{\phi}]$ must be expanded in a series and we must keep the terms up to fourth order in the quantum superfields. Thus, we get 
\bea
S[\Phi+\phi,\bar{\Phi}+\bar{\phi}]&=&\int d^8z\bar{\phi}f(\Box)\phi+\bigg\{\int d^6z\bigg[\frac{1}{2}\phi\left(g(\Box)+W^{\prime\prime}\right)\phi+\frac{1}{3!}W^{\prime\prime\prime}\phi^3\nonumber\\
&+&\frac{1}{4!}W^{\prime\prime\prime\prime}\phi^4\bigg]+h.c.\bigg\} \ .
\eea
It follows from this functional that the propagators are
\bea
\left(\begin{array}{cc}
\displaystyle G_{++}(z,w) & \displaystyle G_{+-}(z,w)\\
\displaystyle G_{-+}(z,w) & \displaystyle G_{--}(z,w)
\end{array}\right)=\left(\begin{array}{cc}
\displaystyle \bar{Y}(\Box)\delta_+(z,w) & \displaystyle F(\Box)\bar{D}^2\delta_-(z,w)\\
\displaystyle F(\Box)D^2\delta_+(z,w) & \displaystyle Y(\Box)\delta_-(z,w)
\end{array}\right) \ ,
\eea
where
\bea
\label{FY}
F(\Box)\equiv-\frac{f(\Box)}{\Box f^2(\Box)-\left|g(\Box)+W^{\prime\prime}\right|^2} \ , \ Y(\Box)\equiv\frac{g(\Box)+W^{\prime\prime}}{\Box f^2(\Box)-\left|g(\Box)+W^{\prime\prime}\right|^2} \ ,
\eea
and the $3$- and $4$-point chiral vertex functions are given by
\bea
\label{4d3vertex}
\frac{\delta^3 S}{\delta\phi(z_1)\delta\phi(z_2)\delta\phi(z_3)}&=&W^{\prime\prime\prime}(z_3)\delta_+(z_1,z_2)\delta_+(z_2,z_3) \ ;\\
\label{3d4vertexa}
\frac{\delta^4 S}{\delta\phi(z_1)\delta\phi(z_2)\delta\phi(z_3)\delta\phi(z_4)}&=&W^{\prime\prime\prime\prime}(z_4)\delta_+(z_1,z_2)\delta_+(z_2,z_3)\delta_+(z_3,z_4) \ ,
\eea
where $\delta_+(z_1,z_2)\equiv\bar{D}^2_{1}\delta^8(z_1-z_2)$ is the chiral delta-function \cite{BK}. Of course, there are also corresponding antichiral vertex functions.

The topologies of the two-loop supergraphs which, in principle, would contribute to the EP are the same as those studied in the previous section, namely, "$\infty$" and "$\ominus$" topologies. However, the only non-vanishing contribution is the following one:
\bea
\Gamma^{(2)}_\ominus &=&\frac{1}{6}\int d^6\bar{z}_1d^6\bar{z}_2d^6\bar{z}_3d^6z_4d^6z_5d^6z_6\frac{\delta^3 S}{\delta\bar{\phi}(z_1)\delta\bar{\phi}(z_2)\delta\bar{\phi}(z_3)}\frac{\delta^3 S}{\delta\phi(z_4)\delta\phi(z_5)\delta\phi(z_6)}\nonumber\\
&\times& G_{-+}(z_1,z_4)G_{-+}(z_2,z_5)G_{-+}(z_3,z_6) \ .
\eea
Substituting the propagators and vertex functions into this expression and using the delta functions to perform four integrals, we find
\bea
\label{4dominus}
\Gamma^{(2)}_\ominus&=&\frac{1}{6}\int d^6\bar{z}_1d^6z_4\bar{W}^{\prime\prime\prime}(z_1)W^{\prime\prime\prime}(z_4)G^3_{-+}(z_1,z_4)\nonumber\\
&=&\frac{1}{6}\int d^8z_1d^8z_4\bar{W}^{\prime\prime\prime}(z_1)W^{\prime\prime\prime}(z_4)F(\Box_1)\delta^8(z_1-z_4)\left[F(\Box_1)D_1^2\bar{D}_1^2\delta^8(z_1-z_4)\right]^2 \ ,
\eea
where we have used four spinor derivatives contained in one of the propagators to rewrite the antichiral and chiral integrals as integrals over full superspace.

Now, with the help of the well-known relation \cite{GGRS}
\bea
\delta^4(\theta_1-\theta_4)D_1^2\bar{D}_1^2\delta^8(z_1-z_4)=\delta^8(z_1-z_4) \ ,
\eea
we can reduce (\ref{4dominus}) to an expression that is local in $\theta$. Therefore, we obtain
\bea
\label{4dominus_2}
\Gamma^{(2)}_\ominus=\frac{1}{6}\int d^4\theta_1d^4x_1d^4x_4\left|W^{\prime\prime\prime}\right|^2\left[F(\Box_1)\delta^4(x_1-x_4)\right]^3 \ .
\eea
This equation leads to integrals which are easy to evaluate if we use the partial fraction decomposition of $F(\Box)$. Thus, it follows from (\ref{4dpolynomial}) and (\ref{FY}) that 
\bea
\label{4dpfr}
F(\Box)=-\frac{a_N^{-1}f(\Box)}{\prod_{k=1}^N\left(\Box-m^2_k\right)}=-a_N^{-1}\sum_{k=1}^N\frac{d_k}{\Box-m_k^2} \ ,
\eea
where the residues are \cite{MH}
\bea
\label{4dresidues}
d_k=\prod_{j\neq k}\frac{f(m_k^2)}{m_k^2-m_j^2} \ .
\eea
Substituting (\ref{4dpfr}) into (\ref{4dominus_2}) and  going to the momentum space, we find
\bea
\Gamma^{(2)}_\ominus&=&\frac{a_N^{-3}}{6}\int d^8z\left|W^{\prime\prime\prime}\right|^2\sum_{k=1}^N\sum_{l=1}^N\sum_{m=1}^Nd_kd_ld_m\int \frac{d^4kd^4l}{(2\pi)^8}\frac{1}{k^2+m^2_k}\frac{1}{l^2+m^2_l}\frac{1}{(k+l)^2+m^2_m} \ .
\eea
Here, we use the analytic formula obtained in \cite{FJJ} to solve these two-loop vacuum integrals with arbitrary masses. Therefore, we obtain
\bea
\label{4dominus_3}
\Gamma^{(2)}_\ominus&=&\frac{a_N^{-3}}{6(16\pi^2)^2}\int d^8z\left|W^{\prime\prime\prime}\right|^2\sum_{k=1}^N\sum_{l=1}^N\sum_{m=1}^Nd_kd_ld_m\bigg\{-\frac{m_k^2+m_l^2+m_m^2}{2\varepsilon^2}-\frac{1}{\varepsilon}\bigg[\frac{3}{2}\big(m_k^2\nonumber\\
&+&m_l^2+m_m^2\big)-L_1\bigg]-\frac{1}{2}\bigg[L_2-6L_1+\left(m_l^2+m_m^2-m_k^2\right)\ln\frac{m_l^2}{\bar{\mu}^2}\ln\frac{m_m^2}{\bar{\mu}^2}+\big(m_m^2+m_k^2\nonumber\\
&-&m_l^2\big)\ln\frac{m_m^2}{\bar{\mu}^2}\ln\frac{m_k^2}{\bar{\mu}^2}+\left(m_k^2+m_l^2-m_m^2\right)\ln\frac{m_k^2}{\bar{\mu}^2}\ln\frac{m_l^2}{\bar{\mu}^2}+\xi(m_k^2,m_l^2,m_m^2)+\big(m_k^2\nonumber\\
&+&m_l^2+m_m^2\big)\left(7+\zeta(2)\right)\bigg]\bigg\} \ ,
\eea
where $\varepsilon=\frac{1}{2}(4-d)$,
\bea
L_n=m_k\left(\ln\frac{m_k^2}{\bar{\mu}^2}\right)^n+m_l\left(\ln\frac{m_l^2}{\bar{\mu}^2}\right)^n+m_m\left(\ln\frac{m_m^2}{\bar{\mu}^2}\right)^n \ ,
\eea
and the definition of the function $\xi(m_k^2,m_l^2,m_m^2)$ is explained in the Appendix.

Once again, the Feynman integrals give rise to spurious divergences. Such spurious divergences (and also some finite terms) are completely cancelled due to the sum of the residues
\bea
\sum_{k=1}^Nd_k=0 \ , \ \textrm{for $N\geq N_f+2$} \ . 
\eea
The validity of this equation was checked with use of Mathematica to be valid at least up to $N=8$.
Therefore, it follows from (\ref{4dominus_3}) that two-loop contribution to the K\"{a}hlerian EP is finite and equal to
\bea
\label{4dominus2loop}
K^{(2)}_\ominus(\Phi,\bar{\Phi})&=&\frac{a_N^{-3}}{3(32\pi^2)^2}\left|W^{\prime\prime\prime}\right|^2\sum_{k=1}^N\sum_{l=1}^N\sum_{m=1}^Nd_kd_ld_m\bigg\{m_k^2\ln\frac{m_l^2}{\bar{\mu}^2}\ln\frac{m_m^2}{\bar{\mu}^2}+m_l^2\ln\frac{m_m^2}{\bar{\mu}^2}\ln\frac{m_k^2}{\bar{\mu}^2}\nonumber\\
&+&m_m^2\ln\frac{m_k^2}{\bar{\mu}^2}\ln\frac{m_l^2}{\bar{\mu}^2}-\xi\left(m_k^2,m_l^2,m_m^2\right)\bigg\}.
\eea
Finally, by substituting the one- and two-loop contributions (\ref{4dfinal1loop}) and (\ref{4dominus2loop}) into (\ref{4ddefeffpot}) we arrive at the full K\"{a}hlerian EP
\bea
\label{4dfull}
K_{eff}(\Phi,\bar{\Phi})&=&\bar{\Phi}\Phi-\frac{1}{32\pi^2}\sum_{k=1}^N m^2_k\ln\left(\frac{m^2_k}{\bar{\mu}^2}\right)+\frac{a_N^{-3}}{3(32\pi^2)^2}\left|W^{\prime\prime\prime}\right|^2\sum_{k=1}^N\sum_{l=1}^N\sum_{m=1}^Nd_kd_ld_m\bigg\{m_k^2\nonumber\\
&\times&\ln\frac{m_l^2}{\bar{\mu}^2}\ln\frac{m_m^2}{\bar{\mu}^2}+m_l^2\ln\frac{m_m^2}{\bar{\mu}^2}\ln\frac{m_k^2}{\bar{\mu}^2}+m_m^2\ln\frac{m_k^2}{\bar{\mu}^2}\ln\frac{m_l^2}{\bar{\mu}^2}-\xi\left(m_k^2,m_l^2,m_m^2\right)\bigg\}.
\eea
Just like the EP in the higher-derivative scalar superfield model, the K\"{a}hlerian EP in the higher-derivative chiral superfield model is also free from divergences for any analytical function $W(\Phi)$.

We finish this section with an application of the result presented above. Let us consider the higher-derivative model studied by us in Ref. \cite{GGNPS}, where the operators are defined as
\bea
f(\Box)=1 \ ; \ g(\Box)=-\frac{\Box}{\Lambda} \ ,
\eea
where $\Lambda>0$. Thus, it follows from this definition that $a_2=\Lambda^{-2}$ and we have to look for the roots of the quadratic equation (see Eq. (\ref{4dpolynomial}))
\bea
-\frac{z^2}{\Lambda^2}+(\Lambda+W^{\prime\prime}+\bar{W}^{\prime\prime})\frac{z}{\Lambda}-|W^{\prime\prime}|^2=0\ .
\eea
The roots of this equation are given by
\bea
\label{4droots}
m_{\pm}^2=\frac{\Lambda}{2}\left[\Lambda+W^{\prime\prime}+\bar{W}^{\prime\prime}\pm\sqrt{(\Lambda+W^{\prime\prime}+\bar{W}^{\prime\prime})^2-4|W^{\prime\prime}|^2}\right] \ .
\eea
Of course, we must make the assumption that $(\Lambda+W^{\prime\prime}+\bar{W}^{\prime\prime})^2>4|W^{\prime\prime}|^2$ in order to guarantee that these roots are different and real.

Inserting $f(m_{\pm}^2)=1$ and (\ref{4droots}) into (\ref{4dresidues}), we get the residues
\bea
\label{4dresidues_2}
d_+=-d_-=\frac{1}{\Lambda\sqrt{(\Lambda+W^{\prime\prime}+\bar{W}^{\prime\prime})^2-4|W^{\prime\prime}|^2}} \ .
\eea
Lastly, substituting (\ref{4droots}) and (\ref{4dresidues_2}) into (\ref{4dfull}) we obtain the full EP
\bea
K_{eff}(\Phi,\bar{\Phi})&=&\bar{\Phi}\Phi-\frac{\Lambda}{64\pi^2}\Bigg[\left(\Lambda+W^{\prime\prime}+\bar{W}^{\prime\prime}+\sqrt{\Delta}\right)\ln\left(\frac{\Lambda+W^{\prime\prime}+\bar{W}^{\prime\prime}+\sqrt{\Delta}}{2\bar{\mu}^2\Lambda^{-1}}\right)+\Big(\Lambda+W^{\prime\prime}\nonumber\\
&+&\bar{W}^{\prime\prime}-\sqrt{\Delta}\Big)\ln\left(\frac{\Lambda+W^{\prime\prime}+\bar{W}^{\prime\prime}-\sqrt{\Delta}}{2\bar{\mu}^2\Lambda^{-1}}\right)\Bigg]-\frac{\Lambda^{6}}{3(32\pi^2)^2}\left|W^{\prime\prime\prime}\right|^2\nonumber\\
&\times&\Bigg\{\left[\ln\left(\frac{\Lambda+W^{\prime\prime}+\bar{W}^{\prime\prime}+\sqrt{\Delta}}{2\bar{\mu}^2\Lambda^{-1}}\right)+\ln\left(\frac{\Lambda+W^{\prime\prime}+\bar{W}^{\prime\prime}-\sqrt{\Delta}}{2\bar{\mu}^2\Lambda^{-1}}\right)\right]^2-\frac{1}{\Lambda^3\sqrt{\Delta^3}}\nonumber\\
&\times&\big[\xi\left(m_+^2,m_+^2,m_+^2\right)-3\xi\left(m_+^2,m_+^2,m_-^2\right)+3\xi\left(m_+^2,m_-^2,m_-^2\right)\nonumber\\
&-&\xi\left(m_-^2,m_-^2,m_-^2\right)\big]\Bigg\} \ ,
\eea
where $\Delta=(\Lambda+W^{\prime\prime}+\bar{W}^{\prime\prime})^2-4|W^{\prime\prime}|^2$. It is worth mentioning that this EP is consistent with the one-loop EP given in \cite{GGNPS}.

\section{Conclusions}

In this paper, we determined the SEP for higher-derivative generalizations of the three-dimensional scalar superfield model and the four-dimensional chiral superfield model. {The new result of our paper in comparison with previous studies of higher-derivative superfield theories, is the explicit calculation of EPs up to the two-loop order in perturbation theory. We showed that the basic strategy to compute EPs in higher-derivative superfield theories is based on decomposition, with the help of the fundamental theorem of algebra, the functional traces and Feynman integrals into well-known traces and integrals of theories without higher derivatives. Accordingly, we encountered divergences at intermediate steps of the calculation. However, since the higher-derivative theories studied here are all-loop finite as it can be proved with use of the superficial degree of divergence, the divergences disappeared from the final results, so that the EPs obtained were finite and exhibited logarithm-like functional structure. Finally, it is worth to point out that our results are quite general due to the fact that they are valid for higher-derivative operators described by polynomials with arbitrary degree.

A possible continuation of this study would consist in, within this methodology developed here, studying two-loop corrections to more sophisticated superfield theories, such as higher-derivative supergauge theories. It is natural to expect that such calculation could be especially interesting from the phenomenological viewpoint. We plan to study in detail this problem in future works.

{\bf Acknowledgments.} The work by A. Yu. P. has been partially supported by the CNPq project No. 301562/2019-9.

\vspace*{4mm}

{\bf APPENDIX}

\vspace*{2mm}

In this Appendix, we provide the definition of the function $\xi(m_1^2,m_2^2,m_3^2)$ \cite{FJJ}:
\begin{itemize}
\item If $2m_1^2m_2^2+2m_1^2m_3^2
+2m_2^2m_3^2\geq m_1^4+m_2^4+m_3^4$, then
\bea
\xi\left(m_1^2,m_2^2,m_3^2\right)&=&4\left(2m_1^2m_2^2+2m_1^2m_3^2
+2m_2^2m_3^2-m_1^4-m_2^4-m_3^4\right)^\frac{1}{2}\Big[L(\theta_1)+L(\theta_2)\nonumber\\
&+&L(\theta_3)-\frac{\pi}{2}\ln2\Big] \ ,
\eea
where $L(x)$ is the Lobachevsky's function, which is defined as
\bea
L(x)=-\int_0^x dt\ln\cos t \ .
\eea
The angles $\theta_i$ are given by
\bea
\theta_i=\tan^{-1}\left[\frac{m_1^2+m_2^2+m_3^2-2m_i^2}{\left(2m_1^2m_2^2+2m_1^2m_3^2
+2m_2^2m_3^2-m_1^4-m_2^4-m_3^4\right)^\frac{1}{2}}\right] \ , \ i=1,2,3 \ .
\eea
\item If $2m_1^2m_2^2+2m_1^2m_3^2
+2m_2^2m_3^2<m_1^4+m_2^4+m_3^4$, then
\bea
\xi\left(m_1^2,m_2^2,m_3^2\right)&=&4\left(m_1^4+m_2^4+m_3^4-2m_1^2m_2^2
-2m_1^2m_3^2-2m_2^2m_3^2\right)^\frac{1}{2}\Big[-M(-\phi_1)\nonumber\\
&+&M(\phi_2)+M(\phi_3)\Big] \ ,
\eea
where
\bea
M(t)=-\int_0^t dx\ln\sinh x \ 
\eea
and $\phi_i$ are given by
\bea
\phi_i=\coth^{-1}\left[\frac{m_1^2+m_2^2+m_3^2-2m_i^2}{\left(m_1^4+m_2^4+m_3^4-2m_1^2m_2^2
-2m_1^2m_3^2-2m_2^2m_3^2\right)^\frac{1}{2}}\right] \ , \ i=1,2,3 \ .
\eea
\end{itemize}

%\vspace{5mm}

\end{document}